\documentclass[10pt,twocolumn,a4paper]{article}

\usepackage[cp1251]{inputenc}
\usepackage{authblk}
\usepackage{url}
\usepackage{amsmath}
\usepackage{amsfonts}
\usepackage{amssymb}
\usepackage{graphicx}
\usepackage{color}
\usepackage{verbatim}
\usepackage{newtxmath,newtxtext}
\usepackage{listings}
\usepackage{siunitx}
\usepackage{xspace}

\usepackage{balance}

\usepackage[colorinlistoftodos]{todonotes}
 \usepackage
  {geometry}
\providecommand{\keywords}[1]
{
  \small	
  \textbf{\textit{Keywords---}} #1
}

\title{Formal Approach for the Verification of Onboard\\ Autonomous
  Functions in Observation Satellites}

\author[2]{Vincent Mussot}
\author[3]{Silvano {Dal Zilio}}
\author[1]{Lo\"{i}c Correnson}
\author[2]{\authorcr Serge Rainjonneau}
\author[2]{Yves Bardout}
\author[2]{Gr\'{e}goire Scano}
\affil[1]{CEA, List, Software Safety and Security Lab, Gif-sur-Yvette, France}
\affil[2]{Institute of Research and Technology (IRT) Saint-Exup\'{e}ry, Toulouse, France}
\affil[3]{LAAS-CNRS, Universit\'{e} de Toulouse, CNRS, Toulouse, France}
\date{}                     
\newcommand{\csm}[1]{{\footnotesize\texttt{#1}}}
\newcommand{\csmblu}[1]{{\color{blue}\csm{#1}}}

\makeatletter
\newcommand{\dotminus}{\mathbin{\text{\@dotminus}}}

\newcommand{\@dotminus}{%
  \ooalign{\hidewidth\raise1ex\hbox{.}\hidewidth\cr$\m@th-$\cr}%
}
\makeatother

\makeatletter
\def \rightarrowfill{\m@th\mathord{\smash-}\mkern-6mu%
  \cleaders\hbox{$\mkern-2mu\mathord{\smash-}\mkern-2mu$}\hfill
  \mkern-6mu\mathord\rightarrow}
\makeatother

\makeatletter
\def \Rightarrowfill{\m@th\mathord{\smash-}\mkern-6mu%
  \cleaders\hbox{$\mkern-2mu\mathord{\smash-}\mkern-2mu$}\hfill
  \mkern-6mu\mathord\Rightarrow}
\makeatother

\makeatletter
\def \rightarrowfill{\m@th\mathord{\smash-}\mkern-6mu%
  \cleaders\hbox{$\mkern-2mu\mathord{\smash-}\mkern-2mu$}\hfill
  \mkern-6mu\mathord\rightarrow}
\makeatother

\makeatletter
\def \Rightarrowfill{\m@th\mathord{\smash=}\mkern-6mu%
  \cleaders\hbox{$\mkern-2mu\mathord{\smash=}\mkern-2mu$}\hfill
  \mkern-6mu\mathord\Rightarrow}
\makeatother

\begin{document}

\maketitle
\sloppy

 \begin{abstract}
   \normalsize
   We propose a new approach for modelling the functional behaviour of
   an Earth observation satellite. We leverage this approach in order
   to develop a safety critical software, a ``telecommand verifier'',
   that is in charge of checking onboard whether a sequence of
   instructions is safe for execution. This new service is needed in
   order to add more autonomy to satellites. To do so, we propose a
   new Domain Specific Modelling Language and the toolchain required for integration into an embedded software. This framework is based on the composition of deterministic finite state
   machines with safety conditions, timeouts, and transitions that
   accept durations as a parameter. It is able to generate
   code in the synchronous programming language Lustre from a
   high-level specification of the satellite. This gives a formal way
   to derive an event-based algorithm simulating the execution of
   telecommand sequence and, thereupon, a provably correct onboard
   verifier.

\end{abstract}

\noindent\keywords{Formal methods, Safety, Autonomous systems, Space  systems}

\section{Introduction}

Over the last decade, autonomy was progressively introduced in most
technological domains. This can be explained by the necessity to add
``autonomous decisions making'' capabilities to implement new
functionalities or to enhance existing ones. However, the architects
of critical systems tend to oppose this trend, which increases the
complexity of validation stages and therefore implies significantly
higher costs to reach the required levels of safety.

In the space industry, most satellites are yet entirely commanded from
the ground, through pre-computed static mission plan which definitely
leaves room for onboard optimization.

Introducing autonomy in observation satellites would improve their flexibility and responsiveness. The associated benefits could range from integrating urgent requests in an ongoing plan, to reducing the memory footprint thanks to a higher compression of cloudy images. This can be achieved by allowing onboard software to modify its own sequence of commands. However, it should be done without impairing its safety. This is why
we carefully consider the validation of an onboard software
able to update the satellite plan by itself.

At present, observation satellites are mostly teleoperated; they
execute sequences of low level, time-tagged {instructions}, named
\emph{Telecommands} (TC), that are generated and verified on the
ground before being uploaded.  These TC sequences are thoroughly
tested to prevent the occurrence of events that may trigger the
fail-safe mechanisms of the satellite. Indeed, any occurrence of such \emph{feared event}
could result in the following chain of actions: interrupting the execution of ongoing mission plan,
shutting down all non essential systems, directing the satellite solar
panels towards the sun and waiting for the satellite to be taken over
by satellite control experts.  This event should be avoided at all
costs, since recovery may take hours to days, during which the mission
of the satellite is interrupted, resulting in substantial shortfall.

A TC sequence can contains hundreds up to thousands of instructions (depending on the covered time span) and can only be uploaded at infrequent intervals (typically four to eight times a day). This situation has several disadvantages. In particular, it makes it impossible to change the satellite plan quickly, for example to add a new, urgent mission element or to react to the detection of clouds that could obscure the ground. In addition, this mode of operation where the whole set of elementary TCs is uploaded involves significant transmission of data at low rate, which prevents the use of small ground stations.

One solution to these problems is to transmit higher level
instructions to the satellite---what we call \emph{Synthetic
  Telecommands} (STC) in Sect.~\ref{sec:satellite}. We can then instruct the flight software to interpret these commands and modify its execution plan accordingly with the proper anticipation time. 
  A key element to implement this new approach is to provide a software toolchain able to ``expand'' an STC into low-level telecommands (with the appropriate time-tags) and then ``merge'' the result into the sequence of instructions that are already planned.

Given the criticality of the application, we seek to formally validate
the algorithms and software used in this process. This is essentially
a multi-constraints problem, since a valid sequence of TC must take
into account strict timing constraints (respect of deadlines);
constraints on the geometry of the satellite (e.g. attitude angles during image acquisition);
priorities assigned to the different missions; constraints on the memory
capacity, etc.

Validating and verifying this kind of software with the reliability
standards and quality level expected from space missions is expensive
and time-consuming. It mainly relies on extensive test campaigns, with
heavy simulations that can only be performed on ground. Throughout the
article, we propose to use a simple software architecture that relies
on a \emph{Telecommand Verifier}. The goal of this small piece of
software is to accept or reject a sequence of telecommands before its
execution. On the satellite processors, we may not rely on extensive
tests and simulation, due to the limited computational power, which is
why a new approach is required for the onboard verification of TC
sequence. However, our approach is not necessarily limited to onboard
verification: it could also replace specific parts of the validation
process performed on the ground, which may result in significant costs
reductions. More globally, one of our goal is to evaluate the use of
formal methods, such as \emph{static analysis tools} and deductive
verification of programs, to validate key elements of the software
architecture of a satellite.

In order to reach the level of confidence required in critical systems, we also need to prove that this verifier is \emph{sound}, meaning that every sequence of TC vetted should be safe for execution by the satellite controller (or at least as safe as a sequence generated on the ground). Lastly, we are also interested by the \emph{completeness} of our verifier, in the sense that it should accept as many sequences as possible.

This approach is comparable to what appears in some software frameworks
that support remote code execution, such as the Java virtual machine~\cite{leroy2002}, where a \emph{bytecode verifier} is in charge
of checking new code before it is executed. Our approach shares the
same advantages. First of all, it reduces the size of the critical
software components that need to be proven correct (the ``trusted
computing base''). Also, it enables a modular approach, since
we can easily change the range and the behaviour of our set of
STCs without the need to modify the rest of our software platform.\\

\noindent\textbf{Contributions and structure of the paper.}
We start by giving a bird's-eye view of the architecture of a Low
Earth Orbit observation satellite, which provides the main target of
our framework. Next, in Sect.~\ref{sec:motiv-our-appr}, we motivate
the need for adding more autonomy and describe our approach for
validating the execution of ``dynamic plans'', directly onboard, by
using a telecommand verifier. Our main contribution is a new method
for deriving this critical software from a high-level description of
the behaviour of a satellite. To this end, we define a dedicated,
formal modelling language, called CSM (see
Sect.~\ref{sec:form-models-comp}), and explain how we can reduce the
problem of accepting a sequence of TC to an \emph{acceptance problem}
(in the sense of formal languages theory) in the CSM model. We have
applied this approach on a realistic space system, that corresponds to
the AGATA technological platform specification, and give a complete
high-level representation of the obtained model in the diagram of
Fig.~\ref{fig:graphComplete}. This model lends itself naturally to an
implementation into a synchronous language, such as
Lustre~\cite{halbwachs1991synchronous}. In Sect.~\ref{sec:framework}
and~\ref{sec:safety}, we show how we can exploit the program resulting
from the compilation of Lustre and discuss the implications on the
safety assessment that can be made to strengthen our claim that the
verifier is sound and valid.

\section{High-Level Description of an Observation Satellite}\label{sec:satellite}

We have experimented our approach with satellites designed for Earth
observation from Low Earth Orbit. The scope of our case study so far
is limited to the satellite model capabilities of the AGATA
technological platform specification~\cite{charmeau2005agata}, which
is able to run decision algorithms and flight software more easily
than a complete satellite simulator. Nevertheless, we designed our
approach with a focus on extensibility and we believe that most of our
work could be transposed to
other kinds of satellites.\\

\begin{figure*}[!h]
  \includegraphics[width=\linewidth]{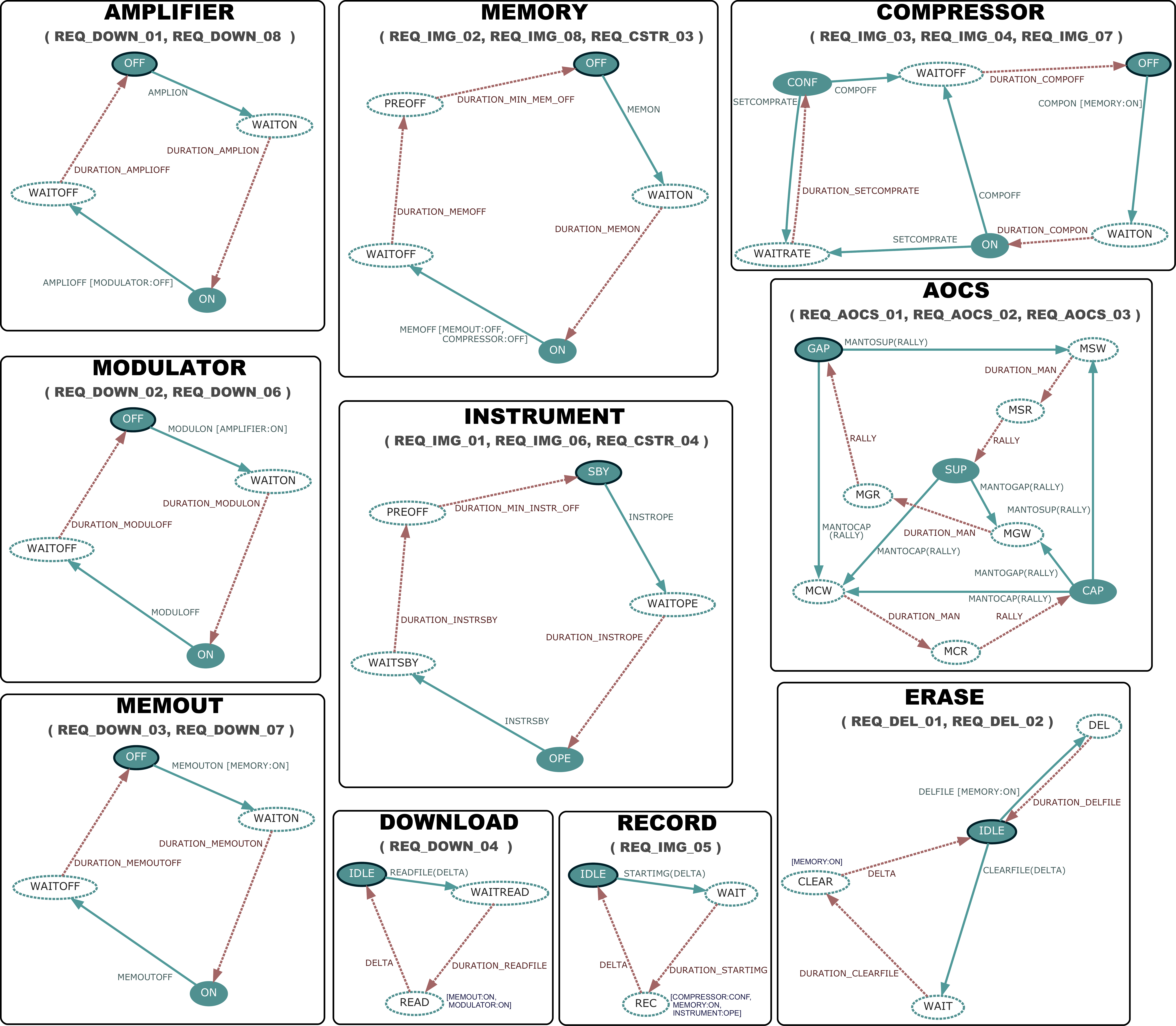}
  \caption{\normalsize Complete graphical model of satellite functions and equipments.}\label{fig:graphComplete}
\end{figure*}

\noindent\textbf{Equipments.} We can describe a satellite based on the set of equipments that it
carries. In our context, the primary mission-related {instrument} is
an optical imager (INSTRUMENT), tasked with capturing images of the
ground. There is also a collection of ``smaller'' equipments, such as
memory banks (MEMORY), a compressor for storage optimisation
(COMPRESSOR), as well as a signal modulator and a signal amplifier,
used for transmitting data to the ground stations (MODULATOR and
AMPLIFIER). All these equipments will appear as separate components in
our
formal model, which is summarized in Fig.~\ref{fig:graphComplete}.\\

\noindent\textbf{Functions.}
Apart from these physical elements, the behaviour of the satellite can
be reduced to the management of several high-level functions. The main
purpose of the observation satellite mission is the \emph{acquisition}
of data (images in our case), using the payload (an optical instrument
in our case). These images stored in the satellite memory must be
collected on the ground, using the \emph{download} function. The
satellite memory can then be freed using the data or file
\emph{deletion} function, to allow the acquisition of new images. Each of these higher-level functions will also correspond
to a component in our model, under the names RECORD (for
acquisition), DOWNLOAD and ERASE.

These functions require the satellite to always be in the right
position and orientation, which is handled by the modes of a
transverse function: the \emph{Attitude and Orbit Control System}
(AOCS). For instance, in the Geocentric Attitude Pointing mode
(\emph{GAP}), the satellite main axis is roughly oriented towards the
Earth center, in order to ease data download. Conversely, during an
acquisition, the AOCS must remain in a Custom Attitude Pointing mode
(\emph{CAP}), to precisely track
the proper footprint on the Earth. The third main mode of the AOCS is
called \emph{SUP} (for SUn Pointing attitude), and is used to point at
the sun and maximize the battery charge from the solar
panels. Each mode will correspond to a state in the AOCS component.\\


\noindent\textbf{Onboard Computer and Telecommands.}
The last component of this architecture is the \emph{onboard satellite
  controller}, which provides the satellite's processing
capability. The controller host the \emph{onboard software} (OBSW)
that is in charge of dispatching orders to the instruments, the
communication between vital functions of the satellites, and the
synchronization and execution of telecommands.

The OBSW can be abstracted as a machine that reads its orders from a
sequence of TC and send resulting commands to the equipments at the
right date. In our context, we can assume that the onboard software is
a periodic task and that each equipment and function can communicate
with the OBSW only at the beginning of each cycle.

Most telecommands are only required to set each equipment of the
satellite in the right mode (typically: switched-on), before
performing a main function. Therefore, in our approach, we consider
only the few high-level requests that are relevant: Acquisition,
Download, Deletion and Maneuver. On this basis, we introduced the
notion of \emph{Synthetic TeleCommands (STCs)}, that can be decomposed
into a sequence of elementary TCs required to perform the corresponding
function.

\begin{figure}[!h]
  \includegraphics[width=\linewidth]{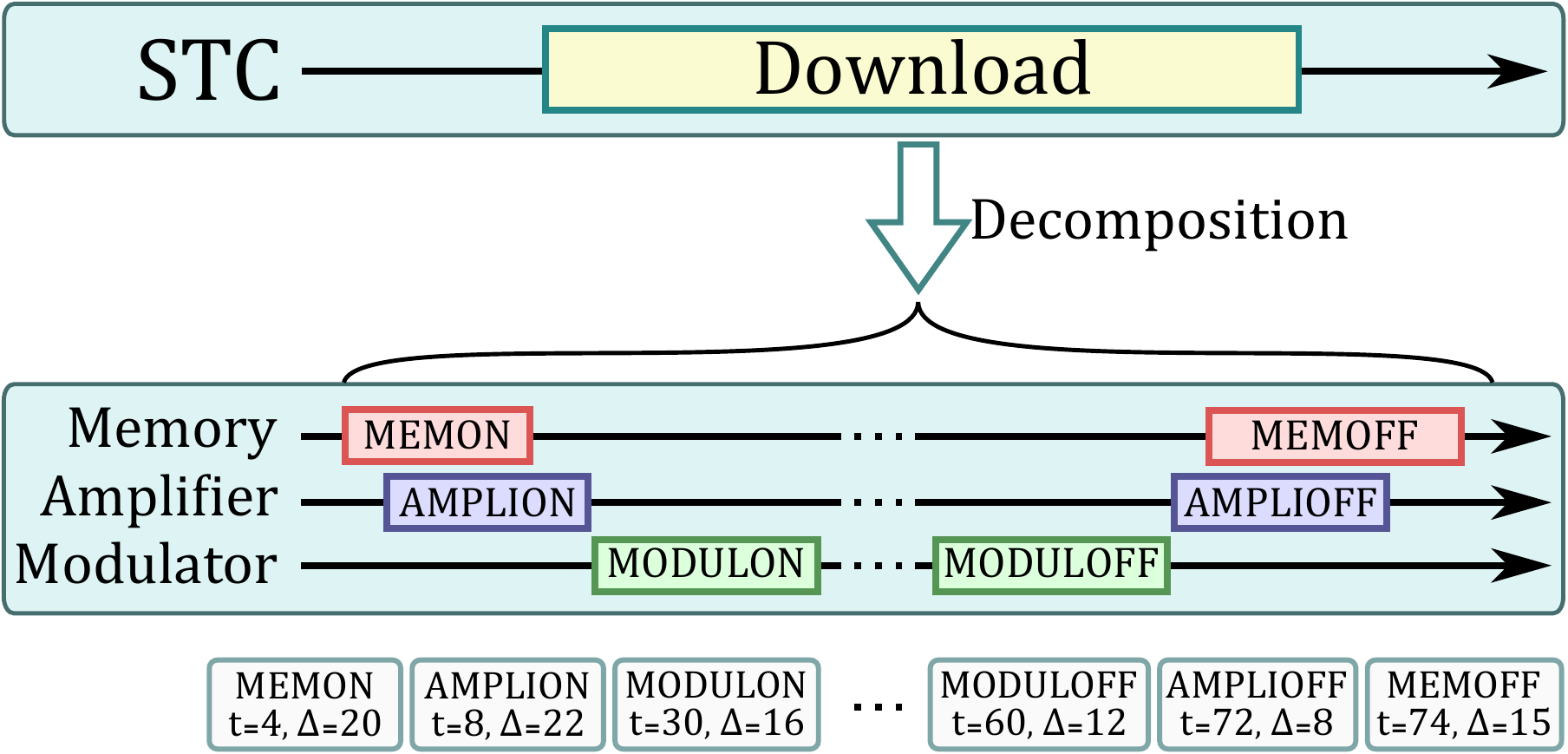}
  \caption{\normalsize Decomposition of an STC into the corresponding TC sequence.}\label{fig:STC}
\end{figure}

We give an example of a sequence of six TCs in Fig.~\ref{fig:STC} that
corresponds to a request (an STC) for downloading an image from memory
to the ground. (In the most general cases, an STC can correspond to up
to eleven TCs.) Essentially, it is a sequence of commands of the kind
\verb+...+$\cdot\,$\verb+MODULON(t=30,+$\Delta$\verb+=16)+$\,\cdot$\verb+...+
with an indication of the absolute date (\verb+t+) and duration
($\Delta$) of each TC. We do not consider other types of parameters
here, such as the memory address where to read data or the position
and the geometry for an image acquisition.

As a first approximation, a sequence of TC is a timed word.
But not all sequences are safe for execution. Indeed, functions and
instruments are closely intertwined together. This gives rise to
several constraints that relate the states of the instruments with the
possible steps of a function. For example, it is unsafe to start
imaging if the satellite is pointed at the sun (when AOCS is in
``state'' \verb+SUP+) since it would expose the optical instrument to
potential damages.

Some functions also introduce timeouts or timing constraints between
the occurrence of events. For instance, transmitting an image requires
to switch the modulator to state \verb+ON+, which takes a specific
time related to physical constraints (defined as a constant
\verb+DURATION_MODULON+ in the
documentation). 

In the following, we develop a formal model that can be used to
describe, in an unambiguous way, whether a sequence of TC is safe. Our
main objective is to derive a \emph{TC verifier} from this
specification. The role of this critical software element is to
reject TC sequences that could harm the satellite's mission.

In the next section, we motivate the notion of STC and explain why the STC decomposition (and the verification of the result) should take place onboard.

\section{Motivation of our Approach}
\label{sec:motiv-our-appr}

The addition of {Synthetic TC} is a new proposal that is motivated as
a way to introduce more autonomy, allowing a satellite to perform some
of its mission planning onboard. It also reduces the amount (and
granularity) of data to be transmitted between the ground and
the board. This extension has an impact on the dependability of the
system. Indeed, the onboard decomposition of an STC is not unique since 
it depends on multiple factors, such as equipments status and variables (e.g. attitude precision), or the  chaining of sequences of STCs. The latter may result in optimization such as keeping an equipment ON if it will be used in consecutive STCs. Therefore, it is not always feasible to test all the possible results of an STC decomposition on the ground, before uploading it for execution.

Actually, our approach is not specifically tailored to the way STCs
are exploited by the onboard software. This allows us to introduce
new extensions incrementally. For instance, in a first class of
autonomy, the satellite may (only) dynamically insert new TCs in its work plan, depending on the STC decomposition strategy. But we
may also envision higher classes of autonomy where a satellite may
choose to schedule or delete an STC dynamically, depending on unpredictable chains of events. For instance, the detection of clouds in an image may lead to its deletion and save memory space, for additional pictures to be captured. Finally, we can even consider cases where
the choice depends on the result of some (black-box) decision
algorithm. A discussion on autonomy in the AGATA platform that we
target can be found in~\cite{pouly2012model}.

The potential optimizations or the ability to consider onboard events
and modify the plan accordingly, when out of reach of a ground
station, are among the strongest arguments in favour of performing the STC decomposition on board. However, its result should be safe for execution by the satellite, to ensure
the continuity of the mission. A way to reduce the complexity of this
task is to verify the sequences of TCs that result from these
decompositions. 
Nevertheless, several issues must be solved in order to
build a TC sequence verifier.

In the following sections, we describe the behaviour of the satellite
using a domain-specific notation and we use this model to define what
is an admissible
sequence of TCs. For the purpose of this work, we compare our
behavioural model of the satellite with a \emph{System
  Requirements Specification} document (SRS) that lists the expected functions
and features related to the operation of the satellite. For instance,
the particular behaviour of the modulator system that we mentioned
above, corresponds to requirement \verb+REQ_DOWN_02+ below, which also
entails that the
signal amplifier must be ON when we power-on the modulator:\\[-2em]
\begin{tabbing}
  \hspace{0.5ex} \= \hspace{2ex} \= \\
   \verb+REQ_DOWN_02+: \textsc{\textsf{Switch modulator to ON}}.\\
   \> \textsf{The modulator is ON after duration} 
  $\Delta$~\verb+DURATION_MODULON+\\
   \> \textsf{initial condition = Amplifier is ON}
\end{tabbing}

Overall, we need to define exactly what it means that a TC
sequence is admissible. This property may depend on the state of the
satellite when the execution starts. Hence we need to define what are
the ``reachable states'' of the satellite, and what is the effect of
executing a command from a state. Finally, we want to provide an
(operational) method to effectively accept or refuse a TC sequence.

We provide a solution to each of these problems. Our main idea is to
use an automata-based framework to describe the behaviour of each
equipment and function. Then the behaviour of the whole system can be
defined as the symbolic composition of all our automata with a
component that represents the onboard software (OBSW). To this end, we
propose a model in which transitions can be triggered by a command
(modelling the reception of an order from the controller) or by a
timeout (modelling the end of an activity). We also provide a way to
associate an invariant to a state (a boolean expression on the states
of the system) that can trigger an error when a condition is not
met. In this context, invariants correspond to error cases identified
in the specification.

\section{Formal Models}
\label{sec:form-models-comp}

In the following, we describe the formal models used in our work. We
provide two different ways (or \emph{notations}) to express the same
model: a graphical notation that is more suited for code review; then
an equivalent textual notation, that is easier to handle with tools,
for code generation and code analysis activities.

We use the textual syntax as our ``pivot'' notation and provide a way
to generate the graphical model automatically from the code in our
toolchain. The main purpose of the graphical model is to ease the
review of the model and the traceability with the SRS.

\subsection{Graphical Notation}\label{sec:automata}

We display the complete graphical model of the satellite in
Fig.~\ref{fig:graphComplete}. Each component (equipment or function)
is named and appears in the form of a finite state machine. We also
list, in each case, the identifiers of the requirements in the SRS
that relates to the behaviour of the component.

We distinguish \emph{steady} states (plain lines)---where we await
orders from the controller---from \emph{transient} states (dashed),
where we await for the end of a timeout. In this case, a timeout is
modelled by an outgoing, dotted transition associated with a
duration. Each component has a unique initial state which is necessarily steady. It is displayed using a thick black border.

Other conventions of our model include: there is always at least one
transient state between two steady states; each transient state is the source of exactly one transition (a timeout). This reflects
some strong invariants in the conception of a satellite: (1) it is not
possible to change state 
without receiving a TC
from the onboard controller; (2) the change of state is
deterministic and the OBSW can always predict which steady state a
TC should switch to (after some finite duration).

If we focus on the automaton for the \emph{Modulator} equipment, we
see that it can be in only two possible steady states: the initial
state, \csm{OFF}, and the \csm{ON} state. Another information given by
the diagram is that only the \csm{MODULON} telecommand can cause a
transition out of the steady state \csm{OFF}; which results in the
Modulator staying in the transient state \csm{WAITON} for a (constant)
duration equal to \csm{DURATION\_MODULON}. This information
corresponds to requirement \verb+REQ_DOWN_02+ of the SRS, that we used
as an example in the previous section.

In order to represent constraints between equipments or functions, we
also define a notation for invariants---displayed inside square
brackets---that can be associated to either telecommands (I) or
specific states (II) as follows:
\begin{itemize}
\item[(I)] Invariants on telecommands are displayed as transition
  \emph{guards} that must be true when the TC is received, otherwise
  the whole TC sequence is rejected (failures are permanent). For
  instance, the telecommand \csm{MODULON} is valid only when AMPLIFIER
  is in state \csm{ON} (during the cycle when the TC was
  dispatched). This is the last element in requirement
  \verb+REQ_DOWN_02+.
\item[(II)] For invariants on states, the implicit behaviour is that a
  component immediately triggers an error when the invariant is
  false. An example can be found on the transient state \csm{READ} of
  the \emph{Download} automaton. Intuitively, this means that the
  \emph{Memory Output} and the \emph{Modulator} should stay \csm{ON}
  the whole time we are downloading an image from the satellite to the
  ground.
\end{itemize} 

One of the reasons that explain the conciseness of this graphical
notation is that we do not need to make explicit many of the ``error
conditions'' of the satellite behaviour. Indeed, we assume an
error when we receive a telecommand that cannot be processed. For
example, it is an error if the \emph{Modulator} receives a
\csm{MODULON} command while in state \csm{ON}. Adding these kinds of
error invariants, for instance by adding a dedicated ``sink state'',
would greatly overload our diagrams.

We can also remark that the OBSW is the only component, defined in
Sect. \ref{sec:satellite}, that is not explicitly represented in the
model. This is because the behaviour of the onboard controller is
fixed. Intuitively, we can think of the OBSW as the component
responsible for dispatching the TC and checking that the invariants
hold.

The resulting model is not far from other timed formal languages,
such as Timed Automata~\cite{alur1994theory} for example. One major
difference, though, is that we may sometimes receive a duration
together with a TC. This is the case in the transition on telecommand
\csm{CLEARFILE} of the \emph{Erase} component/function, originating
from state \csm{IDLE}. This formal parameter, called \csm{DELTA} here,
models the fact that the operation of deleting a file is not performed
in constant time, but depends on its actual size (number of memory
sectors that need to be cleared). In practice, the value of
\csm{DELTA} is part of the TC that is dispatched from the onboard
controller to the \emph{Erase} function and is available to the OBSW
when we check the sequence of TC.


\subsection{Compact Satellite Model}\label{sec:csm}

To simplify our tooling, we designed an equivalent textual syntax; a
programming language called the \emph{Compact Satellite Model}, or CSM
for short. This language can be used to formally describe satellite
requirements. As an example, we provide the CSM specification of the
\emph{Modulator} and \emph{Erase} functions in
Listing~\ref{lst:modulator}.
\begin{lstlisting}[%
  caption={Textual representation of the \texttt{Modulator} equipment and the \texttt{Erase} function in the Compact Satellite Model syntax.},
  float,
  label=lst:modulator,
  %belowskip=-1.5 \baselineskip,
  captionpos=b]
const DURATION_MODULON 2
const DURATION_MODULOFF 3
const DURATION_DELFILE 1
const DURATION_CLEARFILE 1

# REQ_DOWN_02, REQ_DOWN_06
block MODULATOR :=
  init (OFF)
  tc MODULON  (OFF,WAITON,ON) 
              {DURATION_MODULON}
  tc MODULOFF (ON,WAITOFF,OFF)
              {DURATION_MODULOFF}
  guard (MODULON)[AMPLIFIER:ON]

# REQ_DEL_01, REQ_DEL_02
block ERASE :=
  init (IDLE)
  tc  DELFILE (IDLE,DEL,IDLE) 
              {DURATION_DELFILE}
  tcd (DELTA) 
      CLEARFILE (IDLE,WAIT,CLEAR,IDLE) 
                {DURATION_CLEARFILE, DELTA}
  guard (DELFILE) [MEMORY:ON]
  inv (CLEAR) [MEMORY:ON]

\end{lstlisting}

In the CSM, each component definition starts with the keyword
\csmblu{block} followed by the declaration of the initial state---given
after the keyword \csmblu{init}. This state usually represents the
equipment or the function when not in use (\csm{OFF} or \csm{IDLE} in
our running example). The rest of the block is a list of transitions
and invariants declarations.

A transition is defined with the keyword \csmblu{tc} and declares a
sequence, between parenthesis, that starts and ends with a steady
state (the \emph{source} and \emph{destination} states) and that
enumerates all the possible \emph{transient} states in-between, in
order. The delays needed to exit these \emph{transient} states are
given between braces and can refer to durations that are listed in the
SRS (such as \csm{\{DURATION\_MODULON\}} for instance). This notation
makes explicit the constraint that there is at most one sequence of
``transient transitions'' between every pair of steady states (for a
given TC).

We make a distinction between regular transitions, \csmblu{tc}, and
\emph{delta telecommands}, declared with the keyword \csmblu{tcd},
whose timeout depends on a duration that is passed as a parameter of
the TC. We have already described this behaviour in previous section, for the transition \csm{CLEARFILE} of the \emph{Erase} component.

Finally, we use keywords \csmblu{guard} for declaring a guard on
telecommands and \csmblu{inv} for declaring a guard on states, like in
cases (I) and (II) discussed with the graphical notation. Invariants
are the reason why we may not always replace a (deterministic)
sequence of timeouts with a single transition. For instance, if we
look at the sequence of transient states visited after a
\csm{CLEARFILE} telecommand in component \emph{Erase}, we see that
invariant \csm{[MEMORY:ON]} applies in state \csm{CLEAR}, but not in state
\csm{WAIT}. This means that it is not forbidden to request a file deletion when the \emph{Memory} is \csm{OFF}, but is should be \csm{ON}
before the \emph{Erase} function reaches the \csm{CLEAR} state.

Finally, as in the graphical model, the traceability between code and requirements is made simple by the modularity of the language: we can  specify a list of requirements identifier inside the comments of each block (line beginning with the symbol~\csm{\#}).

\subsection{Design Principles for the CSM}

Many of the decisions taken when defining our modelling language are
based on the following design principles:
\begin{enumerate}
\item The model should be readable by any satellite expert that took
  part in writing the satellite specification document, without any
  particular knowledge of formal methods.
\item Building the CSM model from the initial, informal specification
  should not be too tedious or unnecessarily complex.
\item Building the CSM model should ideally not require more time than
  writing the usual spreadsheet, or document in tabular format, used
  to gather system requirements.
\end{enumerate}

The first design principle motivates the choice of an ad-hoc
formalism---a Domain Specific Modelling Language (DSML)---since it
does not require the user to be familiar with a pre-existing
technology.  This helps the verification process, since the formal
model should be reviewed by domain experts, and not software
architects. This also motivates our choice of a notation that is close
to ``state machines'', since the SRS already includes state diagrams in
some of their requirements.

These two design principle help avoid errors and reduce the
time spent by architects on mindless, repetitive tasks. For instance,
the last design principle implies that the CSM model should not
contain more details than the SRS and therefore prevent from  adding non-essential requirements or invariants. These goals are achieved
by designing a language that is tailored towards the exact level of
abstraction needed for the task and that integrates the specificities
of our application domain. For example, we rely on the fact that a
satellite equipment is inherently \emph{time deterministic} (it can
only accept orders coming from the onboard software, OBSW, at a
precise date, and the OBSW can predict in which state every equipment
and function will be in the future). This can also be observed in our
very lightweight treatment of errors.  This would not be possible with
a ``general purpose'' behavioural specification language which, by
definition, requires every details to be made explicit.

There are other incentives for using a dedicated and
``agnostic'' approach, that is unbiased towards any particular
technology. In particular, it may help us take into account changes in
the satellite architecture more easily. Moreover, we could imagine
using the CSM model for something else than deriving our TC verifier;
we mention a possible use for generating covering tests in
Sect.~\ref{sec:safety}. Finally, the choice of a dedicated language
for behavioural specification can shield us from problems that could
arise if we decide to change our target platform. For example, in an
initial prototype of our approach, we experimented with a toolchain
targeting a subset of the C language, the ANSI C Specification
Language (ACSL), and a deductive verification approach based on the
use of Frama-C~\cite{kirchner2015frama}. We eventually decided to select
Lustre for our work, since the generated program invariants (the ACSL
part) were too complex in our use case.

\section{Verification of TC Sequence}\label{sec:seqverif}

The core of our approach relies on the fact that we can express the
problem of accepting a TC sequence as a (timed-word) \emph{acceptance
  problem} on the CSM model. Indeed, a sequence of TC can be
interpreted as an execution trace of the whole model in which all
commands are triggered and no errors are produced. As a consequence, a
TC verifier can be directly derived from this definition by
implementing an interpreter and running it on the TC sequence. In our
work, we define the interpreter using the synchronous language Lustre
and extract an executable from it by compiling the result into C code.


The Compact Satellite Model depicted in Sect.~\ref{sec:csm} does not
support any acceptance function as is. However, most of its design is
based on classical automata formalisms, such as Timed
Automata~\cite{alur1994theory}; the \emph{Discrete Event System}
specification (DEVS) of~\cite{concepcion1988devs}; or the Timed
Transition Systems of~\cite{10.1007/BFb0031995}. For instance, like in
the DEVS model, every state is associated with a given lifespan (which
is infinite in the case of steady states) and there is a single
transition associated with the event of a state reaching its
lifespan. Therefore we can easily define a notion of (timed) trace
acceptance for CSM.

Many specificities of our framework make this definition
simpler. First, every TC targets only one particular equipment or
function (there is no synchronization between components). Also,
interactions between equipments and functions are always mediated by
the OBSW, that operates on a fixed cycle\footnote{The frequency of
  this cycle may be quite low. For instance, in the first generation
  of SPOT satellites, the frequency of the controller is of only
  \SI{8}{\hertz}.}. Consequently, we can reason using a discrete time
model, where each ``tick'' is synchronized with the clock of the
controller. Finally, every timing constraint is punctual and can be
expressed as a number of execution cycles (there is no uncertainty on
the duration of an event and therefore no need to use time intervals).

On the other hand, we need to take into account timeouts situations
(the fact that time elapses) and also the fact that several commands
may be issued at the same date, and therefore that multiple components
may trigger transitions simultaneously. Another major difference with
other popular timed models is that some telecommands take a timeout
(duration) as a parameter. This is the case, for instance, with the
telecommand $\csm{CLEARFILE}$, which takes a formal parameter (called
\csm{DELTA}) in the \csm{tcd} transition of block \csm{ERASE} in
Listing~\ref{lst:modulator}. The value of this parameter, when the TC
is executed, gives the duration that should be spent in state
$\csm{CLEAR}$ before moving to $\csm{IDLE}$ (see
Fig.~\ref{fig:graphComplete}). This means that we may possibly deal
with an infinite number of transitions---one for each value of
\csm{DELTA}---even though, in practice, we could bound the duration
parameters.

For the sake of brevity, we cannot give a complete and precise
presentation of the formal semantics for the CSM language here. Let us
just say that we have defined a structural operational semantics for
the CSM based on a small step reduction relation (in the style
typically used in concurrency semantics) and can define the {meaning}
of a CSM model as a Labelled Transition System (LTS) with discrete
transitions representing the evaluation of telecommands; internal (or
silent) transitions representing local changes in the state of a
component (typically the effect of a timeout); and ``continuous''
transitions representing the passage of time. With this semantics it
is possible to prove that our CSM model of
Fig.~\ref{fig:graphComplete} is ``time-deterministic''. Actually, for
any system that meets the constraints listed in
Sect.~\ref{sec:automata}, we could prove that a given TC sequence
corresponds to at most one path in the LTS; and situations of
deadlocks means that the sequence should be rejected.

In this context, the executions (or traces) accepted by a CSM model
are exactly the finite paths in its labelled transition system,
starting from the initial state. It is not necessary to compute the
whole LTS to test if a given TC sequence is admissible. We only need
to ``execute'' the formal semantics and to check, at each step, that
the invariants described in the CSM model are true. Also, we can
easily extend this notion of acceptance to any given reachable state
in the LTS (instead of only the initial state).

\begin{figure*}[!h]
  \includegraphics[width=\linewidth]{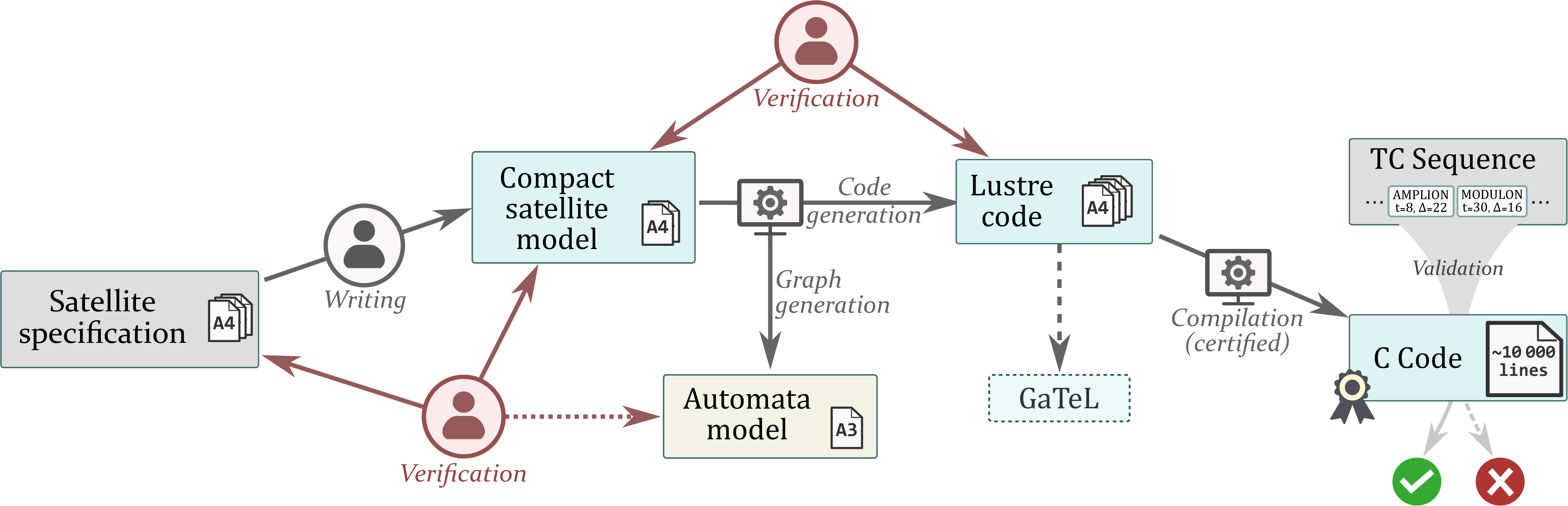}
  \caption{Framework overview.}\label{fig:framework}
\end{figure*}

\subsection{Choices of Operational Semantics}

We can define two main categories of semantics for the CSM. A first
possibility is to define a ``cycle accurate'' interpreter, in which we
simulate the evolution of the system at each cycle. This corresponds
to following each discrete transition in the LTS. If all events are
accepted and constraints between automata are respected, the
simulation continues; otherwise the sequence is rejected and we can
report the last safe state of the system. This approach is easy to
implement, but it may be sub-optimal.

Another possibility is to build a ``discrete-event'' interpreter,
where we can simply skip to the next instant where a meaningful event
occurs (a TC or a timeout). This can speed up the interpreter, but it
requires to build a list combining TC start dates and computed
timeouts, and to sort this list in increasing order.

We have experimented with the two approaches in our work. In each
case, we can provide an encoding of the system model using a
translation into the synchronous programming language Lustre. This
process is made simple by the fact that we can encode each automata
separately and compose them by combining together common events. We
have mostly concentrated our efforts on the ``cycle accurate'' version
of the TC verifier, since it leads to much simpler code and therefore
is simpler to review during the safety assessment.


\subsection{On the Unfolding of STC}
\label{sec:relat-with-unfold}

In this work, we advocate an approach that is somewhat similar to the
one of Proof-Carrying Code (PCC)~\cite{necula1997proof}, a software
mechanism that allows a host system (a satellite in our case) to check
that it is safe to execute a program supplied by an untrusted source.

It would not be realistic to develop a simple, onboard verifier at the
level of synthetic telecommands. Basically, each STC can be compiled
into a fixed sequence of TC. This is the simple part. Then we need to
insert, or \emph{weave}, the resulting sequence into the commands that
are already scheduled by the onboard software. Inserting a TC sequence
into an existing one is a difficult problem when you want to optimize
the result. Indeed, even a slight change in the order of TC can have
far-reaching consequences. For example, delaying the capture of an
image because we have inserted a conflicting command may result in a
longer time spent changing the orientation of the satellite (to
compensate for the change in the direction that the AOCS needs to
point to). Therefore the expansion of an STC is also a scheduling and
planning problem, subject to multi-objective optimization constraints:
we want to acquire a maximum number of usable images; while balancing
the memory allocation; and taking into account the opportunities to
download images that have already been taken. In order to be
competitive with plans computed from the ground---that is to compute a
schedule that is as close as possible to an ``optimal'' one---we need
to use heuristics and techniques from Operational Research. And it
will be very complex to prove the safety of such an STC unfolding and
optimization engine.

In our solution, we separate the safety aspects of the problem from
the computational one. We do not require any knowledge on the way a
new TC schedule is computed from the previous one. We consider the
result TC sequence as an ``untrusted source'', to quote the PCC
approach, and provide a mechanism to check that it is safe for execution.

\section{Overview of our Framework}\label{sec:framework}

We describe our global approach with the diagram in
Fig.~\ref{fig:framework}. The starting point is a set of high-level
requirements (in our example we have chosen the description of a
satellite from the Spot-1 family). These requirements cover the main
equipments of the satellite, as well as the AOCS modes and the
mission-related functions that we described in
Sect.~\ref{sec:satellite}: acquisition, download, memory management,
etc.

The first step consists in deriving a CSM specification from the requirements. Each equipment should typically correspond to a different CSM block and each telecommand can be defined using the keywords \csmblu{tc} and \csmblu{tcd}. The associated durations are also used in these macro commands, as they will correspond to timeouts exiting a transient state.  If an initial condition
must be fulfilled, an \csmblu{inv} or a \csmblu{guard} should be built accordingly, as defined in Sect.~\ref{sec:csm}. 
The resulting model remains very compact. For example, in our use case, the CSM of the whole satellite specification (three pages of requirements) fit into two pages of code.

Once the CSM is obtained, it is used as input by a code generator that
can output both: (1) the graphical model in Graphviz's DOT
language~\cite{ellson2001graphviz};
and (2) a set of components written in the synchronous language Lustre
that implements (time-compatible) ``interpreters'' for each block in
the CSM.

The automata model preserves the modularity and compactness of the CSM
and can be displayed on a single page. It can be used later on to ease
and improve confidence during the verification process between CSM and
specification. Concerning the Lustre code, we give a high-level view
of the structure of the generated code below.

The last step in our framework is to derive the TC verifier by
compiling the Lustre code. For this, we can use one of the many Lustre
compiler that are currently available such as compiler that are
formally verified~\cite{bourke2017formally} or even
certified~\cite{phd/hal/Auger13}. Our objective is to use a certified
toolchain in order to reach the level of safety required in the space
industry and to obtain a C program behaviourally equivalent to the
Lustre code. We review the different parts of this framework again in
Sect.~\ref{sec:safety}, when we discuss safety issues,




\subsection{Structure of the Generated Lustre}\label{sec:lustre}

We generate a set of Lustre nodes (or blocks) that correspond to the
components of the CSM model. Each node can read telecommands by
looking at a specific signals (which are \texttt{on} when the command
is called), and similarly with durations (interpreted as a signal
that is \texttt{on} when a timeout ends).  Every node (CSM component)
can communicate with a central controller, which can be compared with
the OBSW component described in Sect.~\ref{sec:satellite}, in charge
of synchronizing the telecommands and of checking invariants between
components.  The composition of all these components can be used in a
Lustre simulator, such as Luciole~\cite{maraninchi1specification}, to
check the validity of a TC sequence against the specification.


Each node takes as input a dedicated list of telecommand signals and
the corresponding duration parameters if any, and returns its current
state. It handles its own local timers which are usually armed upon
the reception of a telecommand, and is able to modify its return state
accordingly. Finally, it is also responsible for verifying the
compliance between current state and signals received. Each node has a
dedicated error signal, connected to the OBSW node. It will return an
error state if its structural constraints are not respected (for
instance when a node receives a telecommand while not in the expected
state).

The role of the OBSW node is to collect the distributed state of every
equipment and function nodes in order to compute the global view of
the system. This centralized state controller can be used for the
verification of guards and invariants that involve multiple nodes.

The structure of the generated code reflects the modularity of the CSM
specification. This is useful in our case since it means that we can
review the generated code in a modular way; component by component.


\section{Safety Assessment}\label{sec:safety}

This section focuses on the arguments that support our confidence on
the TC verifier obtained with our framework.

One of our main objective is to derive a TC verifier, from a set of
satellite requirements, while both: (1) increasing our confidence on
the tool (for safety reasons); and (2) reducing as much as possible
the need for a human review of the code (for limiting costs and
development time).

To this end, we have designed the CSM language with the goal to limit,
as much as possible, the semantic distance between requirements and
specifications. While we still need to manually review the CSM code to
check its compliance with the requirements, we are able to do this in
only one day in our current use case. This is mainly achieved thanks
to two complementary factors. First, the possibility to automatically
generate a graphical representation from a CSM model\footnote{We could
  increase our confidence by ``certifying'' the generation of the
  graphical model, something that is made easier by the high
  modularity of this transformation.}, which simplifies proofreading
activities.
Second, the possibility to link requirements with transitions in the
CSM model relying on traceability through specific comments. It gives the possibility to perform an analysis of ``model coverage''.

At the other end of the workflow of Fig.~\ref{fig:framework}, the
safety of the compilation step from Lustre to C relies on the use of
certified compilers, and therefore do not need human intervention.

What is left to do is to monitor the transformation step from CSM to
Lustre. For illustrative purpose, in our use case, the Lustre code
derived from our CSM code is only a ``few pages long'' and could be
written by hand in a few days at most. This is to be compared with the
several thousands lines of code in the generated C code.

As described in Sect.~\ref{sec:framework}, we provide an automatic compiler written in OCaml, able to generate the Lustre code from a CSM file. The choice of an intermediate synchronous language preserves the level of abstraction of the CSM, which is by extension the same as the specification itself. Therefore the generated code can be easily reviewed by a Lustre expert since it is only an order of magnitude larger than the original CSM code. Moreover, we can take advantage of the modularity of both languages to perform this verification block per block, and extend the requirements traceability up to this stage.


We are currently exploring ways to increase our confidence on this intermediate step, most of them relying on tests coverage
comparisons. For example, it is possible to generate from the
specification the exhaustive list of minimal TC sequences that lead to
an error, and check the complete coverage of these error cases using
the generated code. It is also possible to generate exhaustive
covering tests directly from the Lustre code, using a tool like
GATeL~\cite{marre2000test}. These tests can then be fed to a
``reference space simulator'', such as the AGATA platform, to
check that a sequence accepted by the verifier cannot trigger a safety
violation in the simulator.

\section{Related Work and Conclusion}\label{sec:relwork}

Autonomy is not a new principle for space systems.  As such, our work
can be viewed as a step forward
in order to increase autonomy at the level of \emph{mission planning
  and execution} (see the various levels of satellite functionalities
classified in~\cite{jonsson2007autonomy}). In particular, we propose a
software architecture able to support the addition of Synthetic TC and
define a cost-effective method for deriving a critical component for
this architecture.

Our approach relies on a new formal model for describing the
functional behaviour of satellites. We show how to leverage this model
in order to develop a safety critical software---a TC verifier---that
is in charge of checking, onboard, whether a sequence of instructions
is safe for execution. We also show how this verifier can be used to
increase autonomy without sacrificing safety.

Our modelling framework is based on the composition of deterministic
finite state machines extended with safety conditions and
timeouts. This is close, in spirit, to several other formal models,
such as the \emph{Discrete Event System} specification (DEVS)
of~\cite{concepcion1988devs} or Timed
Automata~\cite{alur1994theory}. Moreover, our model lends itself well
to a ``compilation'' into Lustre. Therefore, we may have the
possibility to reuse some existing formal verification methods and
tools (such as model simulation or automatic test generation) and
adapt them to our needs. However, our formal modelling language is
also interesting in its own right, since it encompasses many of the
``good practices'' found in space systems---such as time determinism---and enforces them in the
form of syntactical constraints.


\balance
For future work, we expect to extend our approach to other
classes of satellites and to apply it to other problems. For instance,
we would like to reuse our formal models to generate more test cases,
with a better coverage, when testing new onboard planning algorithms.
Also, in the context of safety assessment (see
Sect.~\ref{sec:safety}), our toolchain still requires human review for
several artefacts, and in particular for the Lustre node generated
from the CSM blocks. Automatic verification of the transformation from
CSM to Lustre, using formal techniques, is out of the scope of our
work at the moment. Nonetheless, we could imagine adapting techniques
used in the formal verification of model-based
transformation~\cite{amrani:hal-01083759} to prove that we preserve
the semantics of our models in the compilation. We also mention how we
could use automatic test generation tools, like GATeL, in order to
gain more trust on this step.













\bibliographystyle{plain}
\bibliography{main}

\end{document}